# Review, Meta-Taxonomy, and Use Cases of Cyberattack Taxonomies of Manufacturing Cybersecurity Threat Attributes and Countermeasures

Md Habibor Rahman[a], Thorsten Wuest[b], Mohammed Shafae[a,1]

[a] Department of Systems and Industrial Engineering, The University of Arizona, Tucson, AZ 85721, USA
[b] Department of Industrial & Management Systems Engineering, West Virginia University, Morgantown, WV 26506, USA

## Abstract

The threat of cyberattacks on smart manufacturing systems has been rapidly growing with the potential for a multitude of different attack types, varying from traditional espionage to sabotaging physical assets and products. A thorough and systematic understanding of the different elements of cyberattacks, from motivation to potential consequences and respective countermeasures, is a crucial stepping-stone towards proactive management of manufacturing cybersecurity risks. This understanding is essential for developing the necessary tools to identify, prevent, detect, diagnose, and mitigate cyberattacks. In response, several attack taxonomies have been proposed in the literature as methods for recognizing and categorizing various attributes of cyberattacks, including potential attack vectors/methods, targets/locations, and consequences. However, those taxonomies only cover selected attack attributes depending on the research focus, sometimes accompanied by inconsistent naming and definitions. These seemingly different taxonomies often overlap and can complement each other to create a comprehensive knowledge base of cyberattack attributes that is currently missing in the literature. Additionally, there is a missing link from creating structured knowledge by using a taxonomy to applying this structure for cybersecurity tools development and aiding practitioners in using it. To tackle these challenges, first, this article reviews and analyzes current taxonomical classifications of manufacturing cybersecurity threat attributes and countermeasures, as well as the proliferation of the scope and coverage in current taxonomies. As a result, these taxonomies are compiled into a more comprehensive and consistent meta-taxonomy for the smart manufacturing space. The resulting meta-taxonomy provides a holistic analysis of current taxonomies and integrates them into a unified structure. Based on this structure, this paper identifies gaps in current attack taxonomies and provides directions for future improvements. Finally, the paper introduces potential use cases for attack taxonomies in smart manufacturing systems for assessing security threats and their associated risks, devising risk mitigation strategies, and informing the application of cybersecurity frameworks.

## 1. Introduction

The convergence of Information Technology (IT) and Operational Technology (OT) has transformed once isolated manufacturing systems into data-driven and interconnected smart manufacturing systems. Integrating emerging digital technologies with physical manufacturing processes has created unprecedented opportunities for production automation, real-time data-driven operations, adaptive decision-making and control, and enhanced system visibility [1,2]. However, manufacturing operations are now subject to the same or even more significant cyber threats compared to only IT systems. This is due to the interdependence between cyber and physical assets of the system, the abundance of legacy systems run by outdated software, which is not supported by the providers anymore, and the open-by-design communication protocols to accommodate the heterogeneity of manufacturing assets. Additionally, higher production uncertainty and large data volumes arising from higher product mixes, potential insider threats, complex global supply chains, and unique attack impacts make generic cyber-physical system security methods and guidelines ineffective for smart manufacturing system security. Those factors also make manufacturing organizations lack the visibility required for effective threat detection and response – specifically for OT assets. Previously isolated OT devices – neither designed with cybersecurity in mind nor required before – are now cyber-accessible and have become part of the growing and diverse attack surface.

Increased accessibility, along with the existing and emerging vulnerabilities in industrial control systems and digital manufacturing technologies, are escalating the risk and impact of cyberattacks on manufacturing [3,4]. This is evident in several recent market analysis surveys and reports by cybersecurity solutions providers. For example, as shown in Figure 1, IBM Security reported that the ever-increasing cybersecurity threat to the critical manufacturing industry is now at an all-time high, and manufacturing became the top-attacked industry in 2021 after encountering 23.2% of all cyberattacks [5]. In 2022, the manufacturing industry experienced 72% of all ransomware attacks, when 437 manufacturing entities across 104 manufacturing subsectors (e.g., metal products, automotive, semiconductors, and machinery) were targeted [6]. Moreover, small and medium enterprises, with little to no cybersecurity awareness and limited resources, are attractive targets for cybercriminals. This can have severe impacts on the economy knowing that 60% of small enterprises suffering a data breach close their businesses within six months [7]. Acknowledging this growing risk and the significant need to address manufacturing-specific cyberattacks, the US Department of Homeland Security (DHS) designated the critical manufacturing sector in 2016 as one of sixteen critical infrastructures that "are so vital to the United States that their incapacity or destruction would have a debilitating impact on security, national economic security, national public health or safety, or any combination thereof" [8]. Different government agencies and organizations worldwide (e.g., the European Union, the United Kingdom, and Canada) also acknowledge that adversaries are increasingly targeting manufacturing and industrial processes, and emphasize developing cross-functional knowledge on IT and OT security, enhancing cybersecurity in smart manufacturing systems, incorporating cyber resilience in the product lifecycle, and fostering security awareness [9–11].

---

[1] Corresponding Author.
*E-mail addresses:* habiborrahman@arizona.edu (M.H. Rahman), thwuest@mail.wvu.edu (T. Wuest), and shafae1@arizona.edu (M. Shafae).

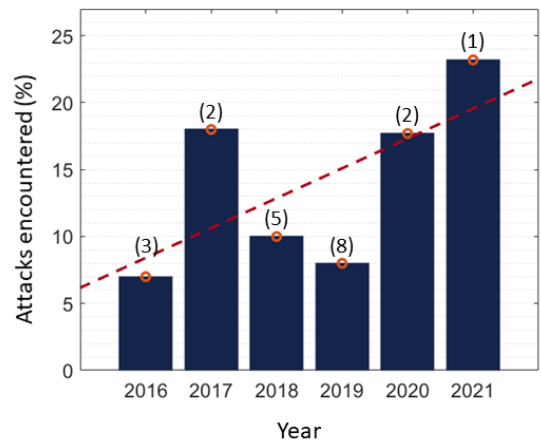

**Figure 1:** Percentage of cyberattacks encountered by the manufacturing industry according to the IBM Security X-Force Threat Intelligence Index, ranks among the top ten attacked industries is shown in parentheses [5,12–15]

Considering the cyber-physical nature of smart manufacturing systems, as opposed to traditional IT systems, cyberattacks against manufacturing be categorized by the targeted/influenced domain and the domain that is affected/victimized, resulting in four attack classes: Cyber-to-Cyber (C2C), Physical-to-Physical (P2P), Cyber-to-Physical (C2P), and Physical-to-Cyber (P2C) [16]. Cyberattacks on manufacturing systems can also be a combination of these classes. For example, a C2C attack can be executed to access the G-code file [2], followed by a C2P attack altering critical process parameters (such as nozzle movement or extrusion rate) in this file (cyber-domain is the target of the C2P attack) to alter with the part geometry (physical-domain is the victim of the C2P attack). Several national institutes and organizations, such as the US Cybersecurity Manufacturing Innovation Institute (CyManII), project that manufacturing cyberattacks will concentrate more on advanced persistent threats or sabotage, rather than solely traditional cyber espionage such as intellectual property theft or ransomware, aiming to disrupt critical manufacturing ecosystems [17]. Those attacks can target/victimize the manufacturing physical domain, including the manufactured products [3,4,18–21], the manufacturing equipment [22–24], and/or the integrated manufacturing ecosystem and sub-systems [25–28]. Multiple academic studies have already demonstrated product-oriented attacks, such as the insertion of voids in a component in stress concentration areas in additive manufacturing or slightly altering the part's geometric integrity, which can result in a significant loss of the part's strength and functional performance [18,19]. An example of a cyberattack targeting manufacturing equipment was on the powder delivery system for sabotaging a Powder Bed Fusion (PBF) additive manufacturing process [22]. Another real-world example of an attack on manufacturing equipment was when adversaries struck a German steel mill, which caused physical components failures in a blast furnace, resulting in massive system damage [23]. Finally, several attacks on manufacturing ecosystems and sub-systems happened in recent years. For example, in 2022, one of Toyota's plastic parts and electronic component suppliers was targeted by a cyberattack, leading to suspended operations in

---

[2] Adversaries can use malware and phishing attacks to steal login credentials (cyber-domain is the target of the C2C attack) from employees and leverage the stolen credentials for accessing digital files such as G-codes in a cloud storage (cyber-domain is the victim of the C2C attack) [18,21,57,58].

Toyota's twenty-eight production lines across fourteen plants in Japan and halting one-third of its global production for more than a day [27].

To address the growing manufacturing cyberattack threat and its devastating impacts, manufacturing stakeholders must rethink their threat landscape and adopt a proactive cybersecurity risk management approach to secure critical smart manufacturing systems. Such approach should prioritize cybersecurity control, develop controls for system resilience, and optimize security investment [29]. The first objective in pursuit of proactive risk management encompasses understanding, identifying, and assessing threat actors; methods, locations, and consequences of potential cyberattacks; and available defense measures in manufacturing systems. Taxonomies can help achieve this objective by offering a consistent and structured classification scheme to systematically characterize and categorize these different attributes of cyberattacks. Recognizing the value of taxonomies, the manufacturing research community has proposed several taxonomies providing taxonomical classifications of the attributes and countermeasures of manufacturing-specific cyberattacks. Several research articles also proposed classifications of specific attributes of attacks in their works which can be considered taxonomical in nature. However, the proposed taxonomies vary due to differences in the scope (e.g., covered attack attributes), level of detail, and organizing principles [30]. As different taxonomies often complement each other, a holistic analysis of the existing taxonomies and their integration under a unified structure enables identifying new research opportunities for characterizing and addressing cybersecurity issues in manufacturing. Such a comprehensive overview of research efforts related to manufacturing cyberattack taxonomies is currently missing. Additionally, while cyberattack taxonomies are valuable in enabling a comprehensive and consistent understanding of cyberattacks, it is not always clear whether and how such taxonomies' enabled understanding can be leveraged to aid the development of different cybersecurity tools in applications such as risk assessment and mitigation and the implications of this to practitioners.

In response to these two gaps, this work presents the first review of manufacturing-specific attack taxonomies and synthesize them to create a meta-taxonomy of manufacturing cybersecurity threat attributes and countermeasures, followed by introducing potential use cases of manufacturing cyberattack taxonomies. First, in Section 2, current manufacturing cyberattack taxonomies are reviewed – showing how the scope and coverage of the taxonomical classifications evolved – and compiled into a meta-taxonomy to offer a unified and consistent picture of different attack attributes. Second, missing attack attributes to realize the complete attack chain and potential extensions of existing taxonomies are pointed out in Section 3. Third, Section 4 provides insights into the value attack taxonomies can offer in assessing cybersecurity threats, as well as their associated manufacturing risks, illustrating how taxonomies can be utilized for risk identification, modeling, and mitigation. Section 4 also discusses the potential use of taxonomies in practice through underlining connections to cybersecurity frameworks, such as the National Institute of Standards and Technology (NIST) framework, highlighting the impact on practitioners and industry. Finally, Section 5 draws the paper to its conclusion.

## 2. Meta-Taxonomy for Cyberattacks in Manufacturing

This section reviews existing manufacturing attack taxonomies and compiles them to create a meta-taxonomy, offering a holistic analysis of the existing taxonomies and integrating them into a unified structure. Previous attack taxonomies to analyze computer and network security-related incidents cannot solely offer a comprehensive grasp of the diversity of possible attacks on cyber-physical systems. Therefore, cyber-physical cross-domain attacks are generally less understood than those in the IT domain [18]. Hence, various research efforts have emphasized developing a common language for understanding potential attacks and defenses specific to manufacturing systems and proposed different taxonomies focusing on 1) attack methods, 2) attack locations, 3) attack consequences, and 4) potential countermeasures. Table 1 presents a brief overview of those taxonomies, summarizing the significant contribution(s) and the progression in the scope and coverage. Some taxonomies provided extensive detail on specific attack attributes, while others overlapped in the proposed classifications. The current taxonomies also may seem different for using inconsistent terminologies and classifications. However, regardless of the differences in scope and classification basis, these taxonomies complement each other in depicting the cyberattack threat landscape. Hence, a comprehensive review of existing taxonomies and their incorporation into a uniform structure assists in characterizing cyberattacks and tackling cybersecurity challenges in manufacturing.

This paper is the first to compile a meta-taxonomy by reviewing, analyzing, and consolidating the current scattered knowledge established by existing attack taxonomies. More specifically, this paper assembles different categories and respective individual elements of attack methods, locations, consequences, and countermeasures presented in current taxonomies into a unified structure. This unified structure is presented in the following subsections, where it groups the main attack attribute classifications proposed in existing taxonomies and provides a brief overview of the attack chain. Section 2.1 summarizes the classifications of attack methods and locations, the two most common dimensions covered in different taxonomies. Section 2.2 and Section 2.3 combine categories of attack consequences and potential countermeasures, respectively, which are needed to inform work on attack prevention, detection, and mitigation.

**Table 1:** Proliferation in the contributions of manufacturing cyberattack taxonomies[3]

| References | Key Contributions | Focus Industry |
|---|---|---|
| Wu and Moon (2017) [31] | Presented physical domain attack vectors and locations for cross-domain attacks on manufacturing systems; illustrated prospective physical consequences of the attacks. | Discrete and continuous-flow manufacturing |
| Pan et al. (2017) [32] | Described potential cyberattacks on IoT-based manufacturing systems; explained cyber-domain vulnerabilities that adversaries can exploit; categorized attack impacts based on confidentiality, availability, and integrity. | Discrete manufacturing |
| Wu and Moon (2018) [33] | Introduced the human element as an attack target and illustrated the associated risks; discussed environmental and operational damages from cyberattacks on manufacturing. | Discrete and continuous-flow manufacturing |
| Tuptuk and Hailes (2018) [34] | Presented manufacturing-specific attack types (such as physical tampering); introduced an adversarial model; categorized potential adversaries; discussed the attack lifecycle. | Discrete manufacturing |
| Wu et al. (2018) [35] | Briefly reviewed different cybersecurity attributes in digital manufacturing, from threat and vulnerability identification to risk assessment; categorized potential countermeasures to detect, counteract, and minimize security risks. | Discrete manufacturing |
| Yampolskiy et al. (2018) [30] | Provided an in-depth discussion on additive manufacturing-specific security threats and prospective attack methods; showed the correlation among attack methods and attack targets; introduced security threats arising from the material supply chain. | Discrete manufacturing focusing on AM |
| Elhabashy et al. (2019) [36] | Expanded the manufacturing cyberattack surface to quality inspection tools; demonstrated the relationships between manufacturing and quality control systems; presented attackers' view through four attack design considerations from the quality control perspective. | Discrete manufacturing |
| Shafae et al. (2019) [19] | Introduced an attack design scheme for describing different elements of cyberattacks on CNC machining systems; classified attack design considerations to subvert detection by traditional quality control tools; and presented how physical attributes of parts can be targeted in product-oriented attacks. | Discrete manufacturing focusing on machining |
| Mahesh et al. (2021) [37] | Classified attack targets into design and manufacturing phases; grouped potential countermeasures against the security threat in manufacturing systems into six categories and associated them with attack targets. | Discrete manufacturing |
| Williams et al. (2023) [38] | Utilized NIST's Cybersecurity Framework to categorize surveyed cyberattacks in the manufacturing sector; attack events were distinguished based on their potential impacts on manufacturing systems. | Discrete and continuous-flow manufacturing |

## 2.1 Taxonomical classifications for attack methods and attack locations

Similar to the Common Attack Pattern Enumeration and Classification (CAPEC[4]) repository [39], a community resource for identifying and understanding attacks, most manufacturing attack taxonomies focus on attack methods and attack locations to distinguish attacks [30–33,36,37]. Different attack taxonomies interchangeably used *attack methods* and *attack vectors* to define *how* an adversary can breach the network/system. Similarly, *attack locations* and *attack targets* were used synonymously to represent *where* adversaries can infiltrate the manufacturing value chain. While most currently available taxonomies covered attack methods and locations, the specific categories they employ vary. We combined the different categories used to date across the taxonomies in Table 2 and Table 3, respectively. The contribution of existing taxonomies to this categorization is explained in more detail in the following subsections.

### 2.1.1 AM specific classification

Focusing on the application in additive manufacturing, Yampolskiy et al. (2018) provided taxonomical classifications of attack targets and methods and showed their correlation [30]. They offered two taxonomies specific to Additive Manufacturing (AM) processes; one was focused on the theft of AM technical data, and the other was on the sabotage of AM assets. Attack targets for theft of technical data included part specification, manufacturing process specification, post-processing specification, and indirect manufacturing. Attack methods were categorized as theft of digital files using malware, reverse engineering through eavesdropping or side-channel analysis[5], and AM equipment analysis to extract process specifications. On the other hand, attack targets for AM sabotage were classified as manufactured parts, AM equipment, and the production environment. Attack methods included tampering with 3D part specifications, attacking the product supply chain, tampering with the AM process via insiders or false data injection, and compromising AM equipment through hardware or software backdoors. This taxonomy categorized attack targets and methods at a granular level – especially providing insights into how threat actors can alter different product features and properties– but only applies to AM processes. The proposed classification helps understand how attacks on AM systems can be launched and distinguish various attack methods.

---

[3] It is worth noting that this table primarily highlights the key unique contribution(s) of each work compared to previously published works, and not necessarily all the contributions of the work.

[4] CAPEC is sponsored by the U.S. Department of Homeland Security (DHS) Cybersecurity and Infrastructure Security Agency (CISA) and managed by the Homeland Security Systems Engineering and Development Institute (HSSEDI) which is operated by The MITRE Corporation (MITRE) [39].

### 2.1.2 Expanded attack surface with quality control systems

The attack surface was further expanded from manufacturing processes to quality inspection systems to illustrate attacks aiming to compromise the quality control tools commonly used in manufacturing systems. For example, Elhabashy et al. (2019) divided the production system into two concurrent sub-systems: the manufacturing system and quality control system, and presented an attack taxonomy to better understand the relationships among Quality Control (QC) systems, manufacturing systems, and cyber-physical attacks [36]. They provided taxonomical classifications of attack methods and attack locations for cyber-physical QC attacks. Attack locations in the inspection system were categorized into two major classes: physical and cyber. Attack methods were classified as altering product design, part quality definitions, and manufacturing processes by tampering with digital files, reporting falsified data through data injection, and acquiring QC implementation data by stealing information. The proposed taxonomy was developed from the attacker's viewpoint to help manufacturers understand how cyber-physical attacks could exploit or misuse QC systems. This is important to consider moving forward from a cybersecurity perspective since QC tools have been an integral part of manufacturing systems and have traditionally been used for detecting out-of-control shifts in products and processes. However, current QC tools are not designed to detect maliciously induced shifts or changes by cyberattacks, and today's QC systems can themselves be compromised during attacks. For example, instead of attacking the product design or production processes, adversaries can alter the Geometric Dimensioning and Tolerancing (GD&T) information in the inspection system, allowing the production of faulty products or discarding conforming products. Hence, considering the QC system as an avenue to expand the manufacturing system attack surface underscores the research and development needs to analyze the vulnerabilities in QC tools, evaluate the effects of attacks on the QC system, and develop new security-aware QC approaches.

### 2.1.3 Attack lifecycle and types of attacks

Depending on the attack motivation and targets, a successful attack may consist of a series of activities from attack launch to execution. Tuptuk and Hailes (2018) explained the lifecycle of cyberattacks and categorized the activities needed during successful attack launches into six phases: 1) initial groundwork for attack design, 2) entry to the targeted system using attack methods, 3) vulnerability exploitation and attack propagation, 4) data sharing and updating instructions (if necessary), 5) carry out attack objectives, and 6) operations to evade detection and extend presence in the system [34]. Acknowledging the lack of understanding of manufacturing-specific vulnerabilities, which could inform how manufacturing systems can be attacked, they classified potential attacks based on their similarity with other networked systems. They also listed thirteen common attack methods for manufacturing systems, considering the general cases of past attacks on other networked systems. Some attacks, such as denial of service attacks, eavesdropping attacks, man-in-the-middle attacks, social engineering attacks, and false data injection attacks, can be launched over network communication systems. Additionally, various malware can be used to launch attacks on manufacturing systems. In contrast, a physical attack, i.e., physical tampering (e.g., de-calibrating a sensor for modifying the input signal), requires access to manufacturing systems equipment, potentially with the help of insiders. The analysis of the attack lifecycle presented in this article demonstrates that threat actors can launch multiple attacks in coordination to achieve their ultimate malicious goals. For example, they can exploit insiders using social engineering attacks to acquire the necessary knowledge about system vulnerabilities and later launch attacks targeting specific vulnerabilities. Thus, analyzing the attack lifecycle can offer insights into which attack locations will be targeted in tandem and by deploying which attack methods. Additionally, the phases of an attack can be factored into designing detection mechanisms and alert correlation.

---

[5] Side-channel analysis refers to analyzing process dynamics variables, also known as process side-channels, such as sound and vibration to reconstruct the 3D object model being manufactured.

### 2.1.4  Attack design elements classification scheme

Researchers also proposed systematic attack classifications to proactively understand how prospective cyber-physical attacks could be designed and implemented to evade traditional quality control techniques such as in-situ process monitoring and post-production metrology. For example, Shafae et al. (2019) proposed an Attack Design & Designation System (ADDS) to systematically categorize and describe the critical elements of potential product-oriented cyberattacks (i.e., attacks tampering with product quality) on machining systems [19]. ADDS was developed based on three critical elements for attack designation: 1) the quality integrity category, which captures the altered physical attributes of the part by the attack; 2) the attack design considerations, which refers to the system information gathered and considered by adversaries to minimize the attack detection likelihood; and 3) the implementation location, i.e., the cyber entity in the manufacturing value chain where the attack is introduced. These critical elements were further sub-categorized, and additional details were provided for defining product-oriented C2P attacks (i.e., attacks aiming to maliciously alter the geometry of parts in production). The quality integrity category was decomposed into geometric quality integrity, surface quality integrity, and material quality integrity. A successful attack can alter the physical characteristics of a machined part within one or more of these three quality integrity categories to change its design intent. The detailed classification of various attack design considerations in ADDS indicates how adversaries might utilize the knowledge of industrial control systems, quality inspection plans, and inspection strategies to tamper with the geometric integrity of parts to evade detection using traditional monitoring techniques. The ADDS classification helped design six different attack scenarios with varying quality integrity targets, attack implementation locations, and necessary design considerations. Studying a variety of attack types with varying designs can help improve the preparedness for potential attacks.

To illustrate, one attack objective can be to increase/decrease the diameter of a particular geometrical feature in the CNC turning of a cylindrical part. This attack on the geometric integrity can be implemented by attacking the machine controller software to shift the workpiece coordinate system from that set by the machine operator. With enough knowledge of the quality inspection system, this attack can evade detection by altering the dimensional characteristics of the part in the automated inspection software to designate altered parts as accepted. Additionally, existing in-situ monitoring strategies to detect conventional causes of variation, such as tool wear and chatter, may not be directly applicable to this product-oriented attack detection. For example, varying the dimension of a local geometrical feature may not significantly alter the features of a measured process variable signal, such as vibration selected to predict tool wear. Hence, monitoring techniques not designed with potential malicious geometrical alterations in mind will fail to detect those alterations. Thus, the taxonomical classifications presented in the ADDS can enable understanding and unfolding of potential attack designs and hence enable pointing out the scope for developing proper defense measures. Similar work is needed to cover other manufacturing processes and equipment as well as other steps in the product realization lifecycle. Such cyber-physical attack design schemes can complement existing cyberattack repositories (such as CAPEC) by enumerating potential cyber-physical attacks to develop a body of knowledge for describing and understanding unique attack designs.

### 2.1.5  Additional attack methods/vectors

Attack methods/vectors reported in different taxonomies on manufacturing cyberattacks are summarized in Table 2, categorized by the domain they target/influence. Although more attack methods were reported in various taxonomies, Table 2 only includes those reported by two or more taxonomies for brevity. Additional reported attack methods include web attacks, cross-site request forgery (CSRF), buffer overflow, zero-day attacks, and replay attacks [31,34]. Note that tampering with digital files is a broader classification encompassing manipulation of data, product specifications, process specifications, product Key Quality Characteristics (KQCs), and GD&T information. It is worth mentioning that social engineering attacks can be launched via cyber means (e.g., spoofing someone to click specific links in spam emails) and physical interactions with employees (e.g., gaining sensitive information through conversations). Hence, social engineering was classified as both a cyber and a physical attack method (C-P). While others categorized it as an attack method [31–33], Tuptuk and Hailes (2018) considered social engineering a pre-entry activity prior to an attack launch [34]. Besides, Wu and Moon (2017) defined collecting and analyzing signals of physical process dynamics variables as eavesdropping attacks [31,33], whereas others defined this attack method as side-channel analysis [30,37]. Considering the primary focus of this paper being analyzing the progression of manufacturing-specific attack taxonomies, the individual attack methods are not described in detail. Nevertheless, Table 2 helps point out papers that discussed different common methods in detail.

### 2.1.6  Additional attack locations/targets

Attack locations/targets reported in different taxonomies on manufacturing cyberattacks are summarized in Table 3. Yampolskiy et al. (2018), Shafae et al. (2019), and Elhabashy et al. (2019) mainly focused on attack locations in the physical domain such as product, machines, production process, inspection system, personnel, supply chain, and production environment [19,30,36]. Additional reported physical domain attack locations are sensors and actuators [33,35,37,40]. Potential attack locations in the cyber domain include operating systems, software, network communication system, and cloud storage [32,33,35,37,40]. Another potential attack location in the cyber domain is the firmware of machines and sensors. While some taxonomies (e.g., [37]) use software and firmware interchangeably and put both under the same category, researchers (e.g., [41]) and practitioners often distinguish between them.

**Table 2:** Examples of attack methods/vectors, reported in different taxonomies, categorized by the domain they target/influence

| Taxonomy | Cyber domain | | | | | | | C-P | Physical domain | | | | |
|---|---|---|---|---|---|---|---|---|---|---|---|---|---|
| | Malware and virus | Denial of service | Eavesdropping | Man-in-the-Middle attack | Software backdoor | False data and code injection | Tampering digital files | Social engineering | Insider | Hardware backdoor | Physical tampering | Reverse engineering | Side-channel analysis |
| Wu and Moon (2017) [31] | ✓ | | ✓ | | | ✓ | | ✓ | | ✓ | | | ✓ |
| Pan et al. (2017) [32] | ✓ | | | | | | | ✓ | | | | | |
| Wu and Moon (2018) [33] | | ✓ | | | ✓ | ✓ | ✓ | ✓ | ✓ | | ✓ | | ✓ |
| Tuptuk and Hailes (2018) [34] | | ✓ | ✓ | ✓ | | ✓ | ✓ | | | | ✓ | | ✓ |
| Wu et al. (2018) [35] | ✓ | ✓ | | | | | ✓ | | | | | | |
| Yampolskiy et al. (2018) [30] | | | ✓ | | ✓ | | ✓ | | ✓ | | | ✓ | ✓ |
| Elhabashy et al. (2019) [36] | | | ✓ | | | ✓ | ✓ | | | | ✓ | | |
| Mahesh et al. (2021) [37] | | ✓ | | | | | ✓ | | ✓ | | | ✓ | ✓ |

**Table 3:** Examples of attack targets/locations, stated in different taxonomies, categorized by the domain they target/influence

| Taxonomy | Cyber domain | | | Physical domain | | | | | | | |
|---|---|---|---|---|---|---|---|---|---|---|---|
| | Operating System and software | Network | Cloud storage | Sensor and actuator | Machines | Product | Personnel | Production process | Operations and supply chain | Inspection system | Production environment |
| Wu and Moon (2017) [31] | ✓ | ✓ | ✓ | ✓ | ✓ | ✓ | ✓ | | | | |
| Pan et al. (2017) [32] | ✓ | | | | ✓ | ✓ | | | | | |
| Wu and Moon (2018) [33] | ✓ | ✓ | | ✓ | ✓ | ✓ | ✓ | | ✓ | ✓ | |
| Wu et al. (2018) [35] | ✓ | ✓ | | | ✓ | | | | | | |
| Yampolskiy et al. (2018) [30] | | | | | ✓ | ✓ | | ✓ | ✓ | | ✓ |
| Shafae et al. (2019) [19] | | | | | ✓ | ✓ | ✓ | ✓ | | ✓ | |
| Elhabashy et al. (2019) [36] | | | | | ✓ | | | ✓ | | ✓ | |
| Mahesh et al. (2021) [37] | ✓ | | | ✓ | ✓ | ✓ | | | | | |

## 2.2 Taxonomical classification for attack consequences

In addition to attack locations and attack methods, researchers categorized the consequences of attacks on smart manufacturing systems as a critical attack attribute and included them in the attack description. However, the classification of attack consequences varies across different taxonomies due to their organizing principle. For example, some focused on translating the Confidentiality, Integrity, and Availability (CIA) triad from cyber domain security to the physical manufacturing domain, while others concentrated on how attacks may affect the NIST's secure manufacturing objectives. The following subsections offer more details on the classifications presented in the manufacturing cyberattack taxonomy literature, and a summary of reported attack consequences is provided in Table 4.

### 2.2.1 The CIA triad-based attack consequences

Pan et al. (2017) proposed a taxonomy classifying different elements of cyber-physical attacks against IoT-based manufacturing processes, with the most detail given to attack consequences [32]. In the proposed taxonomy, attack consequences were classified into three groups: 1) integrity attacks, 2) confidentiality attacks, and 3) availability attacks. More specifically, attacks can affect 1) the confidentiality of digital files; 2) the availability of manufacturing resources, slowing down the production process and/or damaging machines and tools; and 3) the geometric integrity and/or mechanical properties of a part. Integrity attacks were further categorized into material attacks and structure attacks. Material attacks can affect the material strength, color, and surface roughness of the manufactured part, while structure attacks can lead to internal void creation, indents on the surface, and dimensional changes in the part. The categorization of attack consequences in this taxonomy can help rethink the Confidentiality, Integrity, and Availability (CIA) triad of traditional Information Technology (IT) attack consequences in the context of manufacturing systems.

### 2.2.2 Consequences based on the affected domain

Wu and Moon (2017) proposed another classification of cross-domain attacks to help domain experts in the cybersecurity and manufacturing industry understand the nature of attacks in the manufacturing environment [31]. The taxonomical classification of attack consequences was based on the domain being affected due to the attack. They categorized attack consequences into 1) cyber consequences, such as IP theft, privacy leakage, financial fraud, and denial of service; and 2) physical consequences, such as defective product, machine manipulation, malfunction and breakage, loss of system availability, environmental disaster, and risk of death and severe injury. While the classification by Pan et al. (2017) only focused on translating the CIA triad consequences from the cyber domain security to the physical manufacturing domain, Wu and Moon (2017) introduced other consequences to continuous-flow manufacturing processes. The proposed taxonomy introduced the risk to human health and life as a potential attack consequence, which is especially applicable to manufacturing environments with hazardous chemicals and radiation.

### 2.2.3 Consequences leading to environmental and operation damages

In a later publication, Wu and Moon (2018) extended their attack taxonomy and included additional types of cyber-physical attacks, proposing that these attacks can affect humans, products, equipment, intellectual property, the environment, and operations [33]. They also offered a detailed discussion on environmental and operational damages from cyberattacks on manufacturing. Environmental damage was further decomposed into six categories: 1) environmental spills in manufacturing processes such as oil and chemical spills, and biological discharges; 2) increased energy consumption of manufacturing processes; 3) increased emission of pollutants; 4) alterations in the chemical properties of a product; 5) increased energy consumption of individual products; and 6) product's emission during usage. The operational damage was distinguished into four categories: 1) compromised equipment availability, 2) operation schedule change, 3) altered job allocation policy, and 4) manipulated supplier availability; all these can result in significant production delays and reduced machine utilization. This taxonomy can help describe the product and system-level attack consequences in a manufacturing system.

### 2.2.4 Consequences based on NIST's secure manufacturing objectives

Williams et al. (2023) proposed another taxonomy categorizing known cyberattacks based on attack consequences [38]. The proposed taxonomy was driven by NIST's recommended objectives to secure the manufacturing industry. Leveraging the NIST Cybersecurity Framework: Manufacturing Profile [42], they identified five key objectives for defending manufacturing systems that are to maintain: 1) human safety, 2) environmental safety, 3) quality of product, 4) production goals, and 5) trade secrets. These organizational goals can provide insights into how a cybersecurity threat can influence manufacturing systems. With these objectives in mind, they categorized attacks according to their effects on human safety, environmental safety, product quality, production goals, and trade secrets. Within each category, they further decomposed the attack consequences into three groups: 1) low (limited adverse effect), moderate (serious adverse effect), and 3) high (severe or catastrophic effect). The proposed decomposition is subjective and depends on the perceived influence of the attack on products, personnel, manufacturing operations, environment, brand image, finances, and the general public. This taxonomy emphasized how cyberattacks can violate an organization's CIA goals of both information and physical manufacturing systems.

**Table 4:** Examples of attack consequences, discussed in different taxonomies, categorized by the domain they affect/victimize

| Taxonomy | Cyber consequences | | | | Physical consequences | | | | | | |
|---|---|---|---|---|---|---|---|---|---|---|---|
| | IP theft | Privacy/ Confidentiality | Financial fraud | Availability/ DoS | Product quality | Machine breakdown | Operational downtime | Environmental damage | Human safety | Operational delay | Energy consumption |
| Wu and Moon (2017) [31] | ✓ | ✓ | ✓ | ✓ | ✓ | ✓ | ✓ | ✓ | ✓ | | |
| Pan et al. (2017) [32] | ✓ | | | | ✓ | ✓ | ✓ | | | | |
| Wu and Moon (2018) [33] | ✓ | ✓ | | ✓ | ✓ | ✓ | ✓ | | ✓ | ✓ | ✓ |
| Wu et al. (2018) [35] | ✓ | | | | ✓ | | ✓ | | ✓ | | |
| Yampolskiy et al. (2018) [30] | ✓ | | | | ✓ | ✓ | ✓ | ✓ | | | |
| Shafae et al. (2019) [19] | | | | | ✓ | | | | | | |
| Elhabashy et al. (2019) [36] | | | | | ✓ | ✓ | ✓ | | | | |
| Williams et al. (2023) [38] | ✓ | | | | ✓ | | | ✓ | ✓ | ✓ | |

## 2.3 Taxonomical classification for countermeasures

Countermeasures are a technology, process, or action that can prevent or mitigate the consequences of attacks on manufacturing systems. Although current taxonomies did not include countermeasures as a layer or dimension of the taxonomy, we identified relevant articles offering an overview of countermeasures that can be classified into distinct groups which fit well into a taxonomical structure and provides additional value. Some researchers defined countermeasures as potential security controls at a system level, whereas others emphasized designing secured processes based on manufacturing phases in the value chain. The following subsections briefly describe the classifications provided for countermeasures specific to manufacturing.

### 2.3.1 Countermeasures as control methods

Wu et al. (2018) defined countermeasures as control methods for detecting, counteracting, and minimizing security risk and classified those methods. First, they categorized the vulnerabilities into software, hardware, operating system, network, and data. Then, a list of attack targets was presented through which threat actors could exploit different system vulnerabilities. In response to the identified vulnerabilities and threats, they emphasized implementing the following security controls: 1) access control, 2) encryption, 3) authentication, and 4) intrusion detection. Access control, which denotes the implementation of restrictions on access to the cyberinfrastructure and resources, was further divided into role-based, attribute-based, context-based, and view-based access control methods. Encryption is a way to encode data so that only the authorized person can access it. Authentication mechanisms validate a user's credentials before allowing access to digital files. Finally, intrusion detection refers to monitoring computer and network systems for potential security problems. The proposed classification only includes cyber domain countermeasures and falls short of considering non-cyber defenses. While the proposed countermeasures may help protect from tampering and theft of digital files, these might not be sufficient to address the security of typically complicated manufacturing systems of intertwined cyber, physical, and human elements.

### 2.3.2 Countermeasures based on manufacturing phases

Mahesh et al. (2021) recently proposed a threat taxonomy and the corresponding defense measures within the digital manufacturing environment [37]. The attack targets were categorized into two broad categories. Design phase targets included CAD software, Stereolithography file, and G-code. Manufacturing phase targets included machines, sensors, actuators, and controllers. They categorized countermeasures into six categories: 1) watermarking by embedding information in digital files to prove the ownership or authenticity, 2) authentication methods for identity verification, 3) noise injection during production to prevent retrieval of acoustic side-channel information by adversaries, 4) fingerprinting[6] techniques to distinguish manufacturing equipment, 5) obfuscation[7] of digital files to avert reverse engineering of products, and 6) anomaly detection techniques to monitor suspicious system behavior. Additionally, countermeasures were linked to different attack targets showing potential defenses when a specific target in the manufacturing system is under attack. For example, designs or products produced by a specific design file or a machine can be uniquely identified using fingerprinting techniques. Hence, such techniques can be deployed for attacks on design phase targets. However, the proposed taxonomical countermeasure classification might not be generalizable for manufacturing processes other than AM, such as machining. Obfuscation, for example, only applies to AM processes. Additionally, countermeasures such as noise injection during production may be ineffective in machining applications.

---

[6] Similar to the biometric fingerprints, manufacturing equipment also generate distinguishable physical fingerprints, such as sound and thermodynamics response, resulting from system variation, environmental conditions, and manufacturing imperfections. These unique physically unclonable features can be used to register and authenticate production from specific machines.

[7] For obfuscating digital files, multiple segments of QR codes can be embedded in a strategic orientation into the CAD model. The embedded QR codes can prevent reserve engineering of products. Even if threat actors get access to the authentic CAD file, printing the part without knowing the designated slicing orientation will result in printing a faulty QR code, enabling detection of counterfeit parts.

## 3. Discussion on Gaps and Suggested Extensions

This section identifies and discusses several gaps within the existing attack taxonomies and suggests potential extensions towards a comprehensive characterization of attack attributes and devising appropriate countermeasures. A holistic analysis of the existing taxonomies and their integration into the meta-taxonomy helped expose several gaps present in current taxonomies. Collectively, current taxonomies only focused on characterizing attack methods/vectors, attack locations/targets, and potential consequences. However, they overlooked other key threat attributes including threat actors and their intentions and the cyber-physical vulnerabilities of manufacturing systems. Additionally, there is a lack of systematic understanding and categorization of different types of physical deviations that can be induced by cyber-physical attacks (i.e., attacks aiming to inflict physical damage). Moreover, there is limited coverage of potential attack countermeasures in the current literature, and a more comprehensive taxonomical classification of countermeasures is needed. Finally, existing taxonomies categorized attack attributes independently without focusing on their interdependencies. Future works are suggested to focus on formulating and documenting those interdependencies, as well as creating flexible shared knowledge representation methods (e.g., ontologies) to enable efficient attack knowledge documenting, structuring, storage, and retrieval.

The following subsections explain all identified gaps in detail and provide directions for future improvements. Sections 3.1 and 3.3 discuss the rationale for including threat actors and vulnerabilities in manufacturing cybersecurity taxonomies and suggest key elements that must be covered for each attribute, respectively. Section 3.2 delineates the need for taxonomical classifications of specific system/machine behavioral deviations and system damage from attacks. Section 3.4 highlights the deficiency and the future scope in characterizing potential countermeasures against cyberattacks. Finally, Section 3.5 explains the need and a pathway for attack knowledge representation to describe attack attributes and their relationships.

### 3.1 Consideration of threat actors

Current manufacturing cyberattack taxonomies do not analyze and classify threat actors. Systematically understanding the threat actor is crucial in understanding the security risk, adversaries' capabilities, potential attack motives, and strategies, which are all needed in designing robust cybersecurity defenses [43,44]. Threat actors can adopt diverse attack methods based on their past knowledge, access to system data, and available resources. Common threat actors are nation-states, cybercriminals, terrorist groups, rival organizations, hacktivists, and insiders. Understanding the different types of threat actors and their underlying motivations and goals is valuable for determining attack propagation, cybersecurity risk assessment, and forensic analysis after attack incidents. Various threat actors can be identified and distinguished based on observable attack attributes (e.g., methods of operation, targets, and the infrastructure used to execute attacks) and technical indicators (e.g., network signature and domains).

Monitoring and analyzing the attack attributes and network traffic is commonly known as behavioral analysis, while attribution analysis refers to investigating the technical indicators of threat actors. Cybersecurity solution providers develop and deploy various probabilistic models to detect malicious behavior and correlate it to the responsible threat actor [45]. For example, a cybercriminal group named XENOTIME has been identified with capabilities to compromise Industrial Control Systems (ICSs), which was responsible for the 2017 attack on an oil and gas facility in Saudi Arabia [46]. Similar groups have signature attack methods targeting manufacturing companies, the electric power generation sector, nuclear energy, governmental organizations, and defense contractors. Note that identifying threat actors remains challenging, requiring manufacturers to have a combination of technical expertise and access to threat intelligence. When viable, depending on the organizational category and product line (e.g., general-use sheet metal manufacturers vs. defense contractors), manufacturers can narrow down the set of potential threat actors that are most relevant to consider for their specific environment. Appropriate defense measures can then be developed, and resources can be allocated geared towards preventing or mitigating actions of respective threat actors. Analysis and classification of threat actors in relation to other attack attributes can offer a leading edge to manufacturers as devising defense strategies can be better informed by designating expected types of threat actors.

## 3.2 Correlation with attack consequences

Several prior taxonomies provided high-level classifications of attack consequences (e.g., data theft, sabotage, and defective products). Attacks on physical targets can be manifested through physical deviations (consequences) in the system behavior or product characteristics. The deviations induced by such attacks can differ significantly depending on the attack method and the physical target, and there is a lack of systematic understanding and categorization of those deviations in the current taxonomies. Such an understanding can help guide the design of attack prevention and detection methods at the physical layer of the system. For example, the CAD file or the tool path (G-code/CAM file) in a CNC turning operation can be altered to affect the product geometry without affecting the magnitude of the overall power consumption or the machining time [19,20]. However, the local machining times for individual cutting cycles will still reflect the alteration in the product geometry [3]. Developing or modifying in-situ process monitoring techniques that look for such localized changes, even if they are granular) can offer a detective measure against attacks on the geometric integrity of products. Likewise, the knowledge of potential deviations in the process dynamics can enable the detection of attack-induced changes in a process or a product using available physical resources such as in-situ monitoring systems, post-production metrology, and nondestructive testing methods.

## 3.3 Integration of system vulnerabilities

Existing taxonomies overlook system vulnerabilities, the link between attack locations/targets and attack consequences (see Figure 2), which threat actors can exploit during attacks. Vulnerabilities, coupled with the attack strategy, determine the attack pervasiveness, the likelihood of its success, and its impact, eventually delineating the cybersecurity risk. Hence, classification and assessment of manufacturing-specific vulnerabilities encompassing vulnerabilities in the human element, inspection system, and production process are essential to comprehensively analyze attack attributes for risk assessment. It should be noted that identifying and defining manufacturing-specific vulnerabilities should go beyond traditional software and network vulnerabilities. For example, Elhabashy et al. (2020) presented several vulnerabilities in quality inspection systems, such as the improper implementation of QC tools, violation of statistical assumptions of QC tools such as control charts, inadequate data collection for inspection, and inspecting a subset of product features [47]. These vulnerabilities offer novel attack surfaces for threat actors on the one hand, and in contrast, provide new pathways to manufacturers for redesigning traditional QC tools with cyberattack threats in mind. QC tools designed for security will help detect quality and performance deviations/shifts not only due to traditional causes of variation but also due to malicious attacks. Similar works on the human element and production processes are also required, and importantly, manufacturing-specific vulnerabilities should be thoroughly and systematically identified and integrated within attack taxonomies.

## 3.4 Characterization of countermeasures

Being closely related to the attack chain, countermeasures can be incorporated as an additional layer of manufacturing cyberattack taxonomies. Categorization of threat attributes can offer insights into the threat landscape, but threat mitigation requires the characterization of countermeasures. Taxonomical classification of countermeasures can aid organizations in identifying the appropriate countermeasures for their risk profiles and improving the risk posture. Countermeasures can be categorized based on the activity level, deployment location in the digital value chain, and attack response strategy. To illustrate, malicious changes in digital files and process parameters will influence the spindle power in a CNC turning operation, and therefore, in-situ monitoring of the power consumption can help assess the integrity of the machining process and machined parts [3]. This process focused physics-informed attack detection countermeasure can be deployed at the production stage to detect cyberattacks, but not in the product design phase. Additionally, this countermeasure is effective against attacks on product and process integrity but ineffective against attacks aiming to steal IP. In contrast to product and process physics-based approaches, machine learning approaches can be developed for typical cyberattacks (e.g., network intrusion and data tampering) detection leveraging algorithms to identify patterns and anomalies resulting from cyberattacks. Likewise, countermeasures can be specifically designed to prevent specific attacks, lessen the attack probability, and/or respond to an ongoing attack. Different countermeasures have different optimal deployment strategies, require specific extents of coordination with other defenses, and respond to the attack in different ways (e.g., alarms and reports, block/modify specific actions, and system reconfiguration). By understanding the scope, limitations, and constraints in available and potential defense measures, incorporating countermeasures as an additional layer in the taxonomy can ensure seamless integration of threat realization and mitigation.

### 3.5 Attack knowledge representation

In the future, an ontology can be developed to incorporate the proliferation of knowledge on manufacturing-specific vulnerabilities, attack methods, and novel defenses with the current knowledge base established by existing attack taxonomies. Ontologies are expressed in semantic language, allowing data integration across dissimilar data sources. The shared semantics of an ontology will enable integration of the knowledge base produced by attack taxonomies and repositories of known threat patterns, such as CAPEC [39], Common Vulnerabilities and Exposures (CVE) [48], and Common Weakness Enumeration (CWE) [49]. Existing attack taxonomies and repositories are limited to lists and textual descriptions of cyberattack attributes, yielding a rich catalog of these attributes but cannot represent their interdependence and inter-relationships. In contrast, an ontology will help organize domain knowledge and the description of attack attributes and their relationships, which can also be transformed into machine-readable formats [50]. The developed ontology can also provide a mechanism to relate attack incidents and potential responses and improve the cybersecurity situational awareness of manufacturers. An ontology also enables the execution of complex queries and precise searches for efficient knowledge extraction [51]. Hence, it can be used as a recommender system for laying out potential defense measures from search queries specifying the types of manufacturing assets, their location in the manufacturing value chain, and system vulnerabilities. The reasoning component of the ontology allows the deduction of new information and making inferences based on the encoded knowledge, which can enhance attack correlation and detection [52]. However, a thorough conceptual analysis of different attributes of cyberattacks and their interconnections encompassed by various attack scenarios would be required to reach this stage.

## 4. Taxonomy Use Cases in Cybersecurity Tools Development

This section proposes selected use cases depicting how cyberattack taxonomies can be leveraged to aid the research and development of cybersecurity tools in applications such as risk assessment and mitigation (Sections 4.1 and 4.2). Additionally, this section highlights the potential use of taxonomies in practice, through underlining connections to established cybersecurity frameworks such as the National Institute of Standards and Technology (NIST) framework (Section 4.3). While manufacturing attack taxonomies aim to create a common language to characterize cyberattacks and sometimes the respective countermeasures, to-date they are underutilized, if utilized at all, in the research and development of other cybersecurity tools. Furthermore, their usefulness to practitioners is often unclear. As a reminder, taxonomy is the science and practice of classification to categorize information and create sharable knowledge. Importantly, taxonomies can provide a consistent structure to store comprehensible information and provide valuable insights into the field of study [53–55]. Hence, an attack taxonomy can and should serve more purposes than just a list of different attack attributes. In the following subsections, we reflect on those aspects of taxonomy and discuss how taxonomies can be used in and add value to the manufacturing cybersecurity field.

### 4.1 Taxonomy-driven Cybersecurity Risk Modeling and Assessment

An attack taxonomy can be used for the systematic classification and categorization of different attributes of cybersecurity threats. Cybersecurity threats in smart manufacturing systems can be characterized by elements of the attack risk model shown in Figure 2, where threat actors attack specific targets, exploit vulnerabilities through well-designed attack vectors, and produce attack consequences leading to organizational risks. It is worth mentioning that the attack consequences and adversarial objectives and goals do not have a one-to-one match and may differ. Risk models can vary depending on the detail and complexity of threat identification, modeling, and analysis. According to the NIST guidelines for risk assessment, threat events for manufacturing cyberattacks can be characterized by tactics, techniques, and procedures (TTPs) used by threat actors [43]. Attack taxonomies can offer great insights into systematically understanding these TTPs.

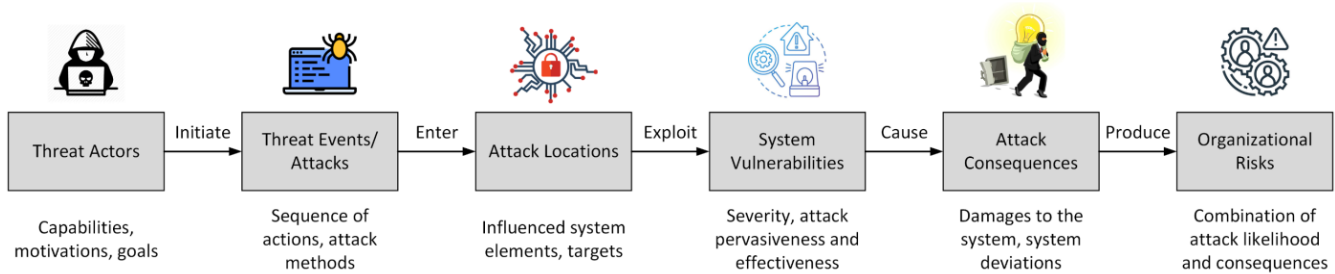

**Figure 2:** A generic risk model for attacks on cyber-physical manufacturing systems

Taxonomies can be utilized to model potential cyberattack propagation paths. The comprehensive structure of a well-developed attack taxonomy can offer insights into a system's vulnerabilities and attack consequences. System vulnerabilities represent potential security states before attacks (pre-conditions), and consequences are related to potential security states after attacks (post-conditions). As opposed to ad-hoc experience-based or example-based models, taxonomy-driven attack graphical models based on pre/post-condition models can enable a comprehensive and consistent realization of attack propagation paths. An example of attack graphical models is the attack trees, a compact graphical depiction of numerous potential attack methods considered as exploit sequences leading to a cybersecurity breach of a system. Attack trees and similar graphical models help assess system weaknesses and enable the planning and deployment of countermeasures. As shown in Figure 2 and described by Tuptuk and Hailes (2018) [34], a successful attack requires a sequence of activities from attack launch to reaching the attack target. Graphical models can offer insights into which activities and sequences can enable attack propagation and determine potential attack paths. Here, graph nodes can represent assets, vulnerabilities, and/or attack targets across the manufacturing value chain. The edges can portray the material and information flow among different nodes. The taxonomical classification of vulnerabilities and attack locations can help populate the nodes of the graphical model, whereas the classification of attack methods can provide the basis to generate the edges. Figure 3 shows a sample attack tree derived from existing taxonomies, depicting potential attack methods and attack paths to reach an attack target. In this example, threat actors can access the network communication system or cloud storage and tamper with digital files (such as CAD models and G-code) to affect the geometry of parts. However, changes in product design would be manifested through changes in process dynamics variables during production and could be detected by sensors. In response, adversaries can inject false sensor data to show that the process operation is nominal. Additionally, they can physically tamper with the production process (e.g., changing machining parameters) for similar malicious intents. The sample attack tree shown here can be extended based on the classifications provided in taxonomies, which can be leveraged to estimate and visualize potential attack paths.

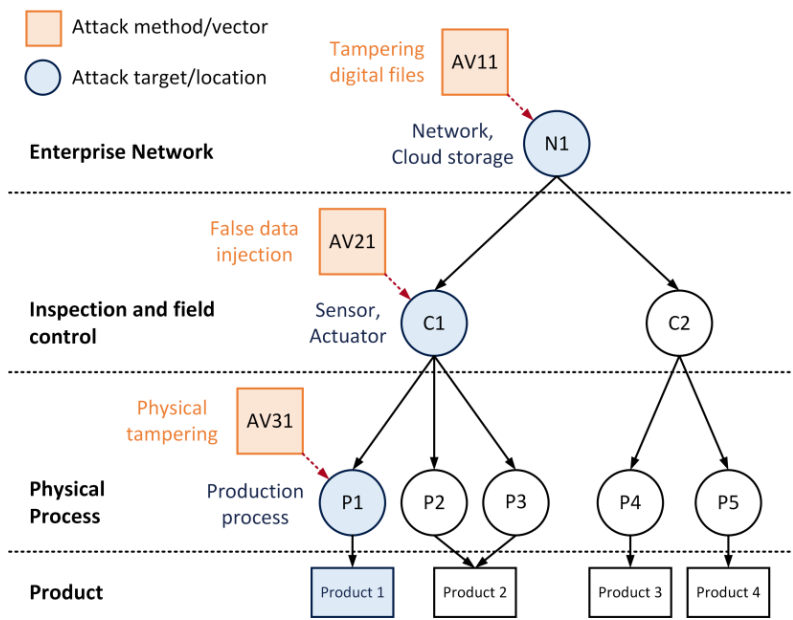

**Figure 3:** Sample graphical model derived from taxonomies, showing attack methods, locations, and propagation

Taxonomy-driven graphical models may also come in handy in developing analytical frameworks for risk assessment. Cybersecurity risk is related to attack pervasiveness and detection probability. Graphical models derived from taxonomies can provide the basis for risk quantification and prioritizing risks. Figure 3 shows that some attack strategies may require attacking several nodes (N1 and C1) in tandem through simultaneous attacks (AV11 and AV21), while other strategies may rely on attacking a single location (P1) and use one attack method (AV31). Such insights can be factored into quantifying an attack's success probability and the likelihood of avoiding detections. For example, coordinated attacks on several locations in the manufacturing value chain may require more resources and increase the detection probability, making the attack costlier for adversaries. Depending on the available resources and adversarial intent, adversaries may have different preferences in selecting the shortest attack path to optimize the damage while minimizing the detection probability. The shortest path for adversaries translates to vulnerable assets and critical connections requiring prioritized defense measures for the manufacturer.

As a proactive security step, such shortest attack paths can be computed using graph-theoretic algorithms such as Dijkstra's algorithm. Similarly, graphs containing system vulnerabilities as nodes and their interconnections as edges can be used to develop vulnerability indices for different manufacturing assets. Whether analyzing attack paths or identifying the most vulnerable manufacturing assets, manufacturing attack taxonomies can serve as the core of generating graphical models and help quantify the cybersecurity risk. The main benefit of developing attack taxonomy-driven graphical models is to ensure comprehensive and consistent coverage of potential nodes and edges of these graphs as opposed to ad-hoc experience-based populated graphs.

## 4.2 Taxonomy-Driven Risk Mitigation

In addition to guiding the cybersecurity risk assessment, taxonomies can help develop proper security solutions by bridging attack consequences to potential risk mitigation techniques. The *threat-oriented* cybersecurity risk assessment approach identifies existing attack cases, finds vulnerabilities from the context of attacks, and identifies consequences based on malicious intents [43]. However, information on the malicious intent may be incomplete, and the performed risk assessment can be limited. In contrast, the *asset/impact-oriented* security risk assessment identifies the consequences of attacks and finds potential attack scenarios that can lead to specific consequences. This approach can help enumerate potential attacks and guide the design and development of possible countermeasures and mitigations by focusing on the inflicted changes in the system. An example of such an approach in discrete part manufacturing is the ADDS, described in Section 2.1.4, which was proposed to systematically design and designate attacks on machining systems with a focus on the physical consequences and mitigations [4]. The detailed classification of various attack design considerations in ADDS provided a systematic scheme to indicate how adversaries might utilize knowledge of human cognitive abilities, industrial control systems operations, traditional process monitoring, and quality control practices to tamper with the geometric integrity of parts while evading detection. Such taxonomical classifications enable the systematic development and study of potential attack designs and the corresponding consequences. This, in turn, allows the research and development of proper countermeasures and mitigations that may not be considered otherwise. For example, by demonstrating various cyberattack-induced localized geometrical changes and how traditional quality control and monitoring techniques fail to detect those changes, Shafae et al. (2019) emphasized the need to rethink in-situ monitoring techniques. This is to allow for detecting small shifts in local amplitude and time signatures of various process variables, which may not be caused by machining traditional causes of variations such as chatter and tool wear.

Attack taxonomies can offer the knowledge of who is carrying out attacks, their intent, and targeting aspects, helping organizations narrow the set of threat events that are most relevant to consider. Recognizing the attack paths can raise awareness about the threat and help manufacturers design efficient protective actions. Referring to Figure 3, once potential attack targets, access points, and attack paths are assessed, appropriate defense measures can be deployed at the proper location. Figure 4 shows potential countermeasures and corresponding implementation locations to secure the attack target based on the example scenario illustrated in Figure 3. For example, threat actors can spoof an employee using social engineering attacks to collect the login credentials and access the cloud storage for CAD files and tamper with those files. In this attack case, watermarking techniques can verify the integrity of the design files upon retrieval at the cloud storage level. At the field control level, fingerprinting techniques can detect malicious data injections, while monitoring process dynamics variables as side channels can detect alternations in production processes at the actual physical process layer. Similarly, other potential attacks and corresponding defense measures can be mapped using attack taxonomy in a comprehensive and consistent manner.

Taxonomy-driven graphical models, such as those discussed in Section 4.1, can also help respond to and recover from cyberattacks. Graph-theoretic approaches, such as the minimum cut, can help isolate the affected nodes (attack locations) during an attack to minimize attack propagation. Attack graphs such as the one shown above can be transformed into Bayesian attack graphs by assigning conditional probabilities to graph nodes. This transformation will enable updating the success probability of an attack as the attack unfolds along the attack path. The Bayesian network structure can also offer a basis for designing an efficient defense scheme for risk mitigation. One example is finding the Pareto-optimal sequences of defense measures deployment that ensures cost-effectiveness while reducing the expected loss from potential attacks.

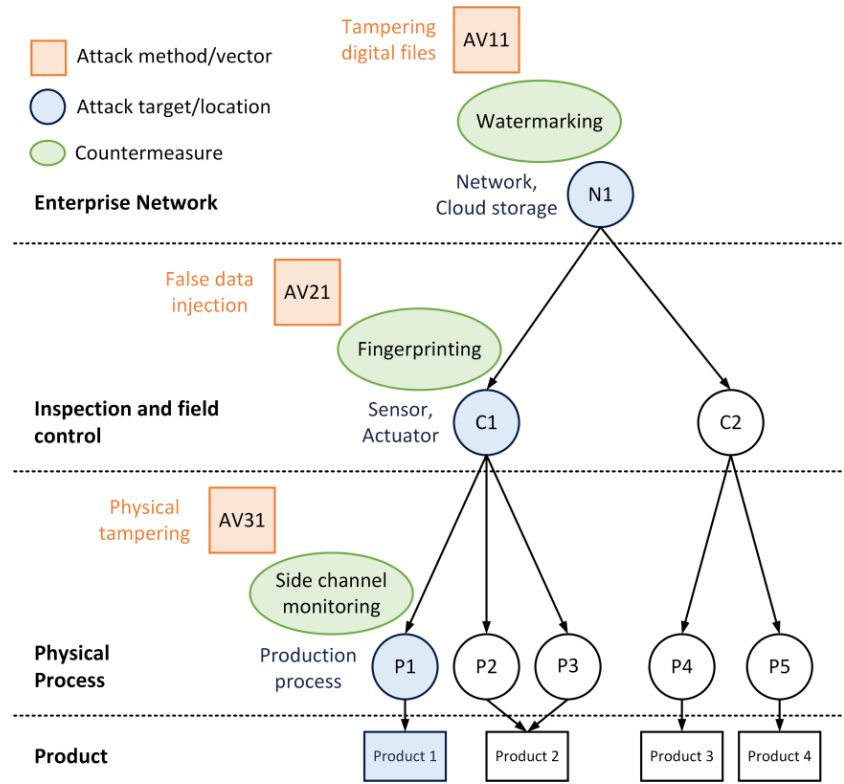

**Figure 4:** Prospective countermeasures can be deployed at different locations within the smart manufacturing system, depending on the perceived risk from potential attack vectors

## 4.3 Taxonomy-Informed Implementation of Cybersecurity Frameworks

Taxonomies can help practitioners interpret and implement cybersecurity frameworks that are designed as integrated approaches to tackle the threat landscape and foster developing capabilities for threat detection and proactive responses. Such frameworks consist of guidelines and best practices for organizations to help them manage cybersecurity risk. For example, the NIST cybersecurity framework was developed to improve cybersecurity risk management in critical infrastructure and strengthen system resilience [56]. In particular, the Manufacturing Profile of this framework, released in 2017, was established as a roadmap to reduce cybersecurity risk for manufacturers [42]. The NIST framework has five core functions: *identify*, *protect*, *detect*, *respond*, and *recover*. These core functions were designed to support enterprises in organizing information, making risk management decisions, and addressing cybersecurity threats. However, the guidelines presented in the framework are typically generic and depend on the manufacturer's risk management process and priorities. As stated in the framework, to effectively manage the cybersecurity risk requires a clear understanding of the business drivers and security considerations specific to the manufacturing system and its particular environment. In response, taxonomies can offer this required understanding through common, consistent, comprehensive, and structured language to categorize, characterize, and communicate cybersecurity considerations such as attack attributes and countermeasures.

To illustrate, the *identify* core of the framework has a category concerning *risk assessment* that requires assessing the risk of manufacturing systems (including threats and vulnerabilities) and attack impacts on manufacturing operations. This category as a guideline is very relevant, but the framework does not provide details on how to conduct the risk assessment. As illustrated in Section 4.1, taxonomies can aid the risk assessment process and support its implementation in a comprehensive and consistent manner. Additionally, a cyberattack taxonomy can serve as the foundation for a standardized and uniform language, helping researchers and practitioners share knowledge and information, fostering collaboration and decision-making, and thus, helping implement the *risk management strategy* category of the *identify* function of the framework. The *protect* core has a category called *awareness and training* that emphasizes providing security awareness training to all manufacturing system managers. In doing so, taxonomies can provide a holistic risk landscape, from threat actors and their malicious intents to potential attack methods and respective countermeasures, which aids in raising security awareness. Likewise, as explained in Section 4.2, taxonomies can be a driving force for developing proper countermeasures and mitigations against potential attacks, supporting the implementation of *anomalies and events* category of the *detect* core that recommends monitoring manufacturing systems to enhance the ability to identify inappropriate activities. The *mitigation* of the *respond* core of the framework suggests mitigating cybersecurity incidents in manufacturing systems and ensuring that vulnerabilities are identified. As discussed before, a well-developed taxonomy can help mitigate cybersecurity threats (Section 4.2), as well as systematically identify associated vulnerabilities (Section 3.3). Additionally, the *recovery planning* category of the recover function of the framework advocates developing and executing recovery plans during or after attack incidents. As explained in Section 4.2, taxonomy-driven graphical models can aid in devising recovery plans during cyberattacks (e.g., isolating parts of the systems in case of a security breach). In summary, the structural framework provided by the taxonomy can help practitioners navigate the implementation of general cybersecurity frameworks like the one proposed by NIST.

## 5. Conclusion

The manufacturing sector has become an attractive target for non-conventional cyberattacks aimed at products, processes, and the entire manufacturing ecosystem, with potentially catastrophic consequences. The growing adoption of digitalization and digital transformation in manufacturing added additional attack vectors to the often already vulnerable IT and OT systems. To defend smart manufacturing systems and protect manufacturers from high-stakes cyberattacks, a systematic understanding of various attack attributes and potential countermeasures is essential. Cyberattack taxonomies address this essential need by systematically and holistically understanding and identifying the methods, locations, and consequences of potential cyberattacks to raise awareness and improve the resiliency of the production system. However, existing taxonomies only partially cover different attack attributes, often with inconstant naming and definitions for the classification. Furthermore, they do not leverage the connection between taxonomy development and its application and implications for cybersecurity tools research and development. This paper presented the first comprehensive overview of recent research efforts in developing manufacturing cyberattack taxonomies, constructed a more comprehensive meta-taxonomy of attack attributes and countermeasures, and proposed several use cases on how taxonomies can be leveraged for assessing security threats and their associated risks, devising risk mitigation strategies, and informing the application of cybersecurity frameworks.

This paper underscored the use and value of cybersecurity taxonomies to understand and characterize cybersecurity threats and attack propagation in manufacturing systems. Additionally, through analyzing current taxonomies and compiling them into a comprehensive and consistent meta-taxonomy, missing attack attributes and scope for future improvements were identified and discussed. For example, countermeasures could be incorporated as an additional layer in attack taxonomies, which can be characterized based on the activity level, deployment location in the digital value chain, and attack response strategy. Connecting system exploits with countermeasures can help manufacturers devise appropriate risk mitigation strategies. Current taxonomies do not account for identifying and analyzing threat actors, which could aid in navigating adversaries' capabilities, potential attack motives, and techniques. Moreover, it is also necessary to link attack outcomes to possible countermeasures and analyze manufacturing-specific system vulnerabilities to fully exploit the potential of an attack taxonomy. Integrating these missing attributes in the future, an attack taxonomy can offer the knowledge of potential adversaries, their intent, and targeting attributes to potential consequences and countermeasures. Future works should also focus on correlating individual attack attributes and modeling those relationships using knowledge representation methods such as ontologies to enable efficient attack knowledge documenting, structuring, storage, retrieval, and inference. Well-developed taxonomies can provide a comprehensive, consistent, and common view of the cybersecurity risk posture to an organization's management, enabling research and development work on making informed risk assessment and mitigation decisions. This work is the first to compile all this information in a single document and should aid researchers and practitioners in understanding what has been studied and what can be further improved in cyberattack taxonomies and potential use cases to enable the design and operation of cyber-physical secure smart manufacturing systems.

## Declaration of Competing Interest

The authors declare that they have no known competing financial interests or personal relationships that could have appeared to influence the work reported in this paper.

## Acknowledgment

The authors would like to express their gratitude for the insightful comments and suggestions provided by Dr. Logan Sturm, Alvin M. Weinberg Distinguished Staff Fellow, Embedded Systems Security Group, Oak Ridge National Laboratory. This research was partially funded by Arizona's Technology and Research Initiative Fund (TRIF) under the National Security

Systems Initiative and the National Science Foundation under Grant No. 2119654. Any opinions, findings, and conclusions or recommendations expressed in this material are those of the author(s) and do not necessarily reflect the views of the National Science Foundation.

## References

[1]  Lu Y, Xu X, Wang L. Smart manufacturing process and system automation–a critical review of the standards and envisioned scenarios. J Manuf Syst 2020;56:312–25. https://doi.org/https://doi.org/10.1016/j.jmsy.2020.06.010.

[2]  Wang L, Torngren M, Onori M. Current status and advancement of cyber-physical systems in manufacturing. J Manuf Syst 2015;37:517–27. https://doi.org/10.1016/j.jmsy.2015.04.008.

[3]  Rahman MH, Shafae M. Physics-based detection of cyber-attacks in manufacturing systems: a machining case study. J Manuf Syst 2022;64:676–83. https://doi.org/10.1016/j.jmsy.2022.04.012.

[4]  DeSmit Z, Elhabashy AE, Wells LJ, Camelio JA. An approach to cyber-physical vulnerability assessment for intelligent manufacturing systems. J Manuf Syst 2017;43:339–51. https://doi.org/https://doi.org/10.1016/j.jmsy.2017.03.004.

[5]  IBM Security X-Force Threat Intelligence Index 2022. https://www.ibm.com/security/data-breach/threat-intelligence/ (accessed December 17, 2022).

[6]  2022 ICS/OT Cybersecurity Year in Review | Dragos 2023. https://www.dragos.com/blog/industry-news/2022-dragos-year-in-review-now-available/ (accessed February 23, 2023).

[7]  Comerford L. Why small businesses are vulnerable to cyberattacks | Security Magazine 2022. https://www.securitymagazine.com/blogs/14-security-blog/post/97694-why-small-businesses-are-vulnerable-to-cyberattacks (accessed August 16, 2022).

[8]  Critical Infrastructure Sectors | Homeland Security 2020. https://www.dhs.gov/cisa/critical-infrastructure-sectors (accessed December 30, 2022).

[9]  Industry 4.0 cybersecurity: challenges & recommendations 2019. https://www.enisa.europa.eu/publications/industry-4-0-cybersecurity-challenges-and-recommendations (accessed February 15, 2023).

[10]  National Cyber Threat Assessment 2023-2024 - Canadian Centre for Cyber Security 2022. https://cyber.gc.ca/en/guidance/national-cyber-threat-assessment-2023-2024 (accessed February 15, 2023).

[11]  National Cyber Strategy 2022 - GOV.UK 2022. https://www.gov.uk/government/publications/national-cyber-strategy-2022/national-cyber-security-strategy-2022 (accessed February 15, 2023).

[12]  IBM Security X-Force Threat Intelligence Index 2017. https://securityintelligence.com/ibm-x-force-threat-intelligence-index-2017/ (accessed March 8, 2022).

[13]  IBM Security X-Force Threat Intelligence Index 2018. https://securityintelligence.com/2018-ibm-x-force-report-shellshock-fades-gozi-rises-and-insider-threats-soar/?mhsrc=ibmsearch_a&mhq=x-force threat intelligence index 2018 (accessed March 8, 2022).

[14]  IBM Security X-Force Threat Intelligence Index 2019. https://newsroom.ibm.com/2019-02-26-IBM-X-Force-Report-Ransomware-Doesnt-Pay-in-2018-as-Cybercriminals-Turn-to-Cryptojacking-for-Profit?mhsrc=ibmsearch_a&mhq=x-force threat intelligence index 2019.

[15]  IBM Security X-Force Threat Intelligence Index 2020. https://securityintelligence.com/posts/threat-actors-targeted-industries-2020-finance-manufacturing-energy/ (accessed March 8, 2022).

[16]  Yampolskiy M, Horvath P, Koutsoukos XD, Xue Y, Sztipanovits J. Taxonomy for description of cross-domain attacks on CPS. Proc. 2nd ACM Int. Conf. High Confid. networked Syst. - HiCoNS '13, 2013, p. 135–42. https://doi.org/10.1145/2461446.2461465.

[17]  CyManII Roadmap 2022. https://www.energy.gov/eere/articles/does-cybersecurity-manufacturing-innovation-institute-releases-first-public-roadmap (accessed December 20, 2022).

[18]  Sturm LD, Williams CB, Camelio JA, White J, Parker R. Cyber-physical vulnerabilities in additive manufacturing systems: A case study attack on the. STL file with human subjects. J Manuf Syst 2017;44:154–64. https://doi.org/https://doi.org/10.1016/j.jmsy.2017.05.007.

[19]  Shafae MS, Wells LJ, Purdy GT. Defending against product-oriented cyber-physical attacks on machining systems. Int J Adv Manuf Technol 2019:1–21. https://doi.org/10.1007/s00170-019-03805-z.

[20]  Wells LJ, Camelio JA, Williams CB, White J. Cyber-physical security challenges in manufacturing systems. Manuf Lett 2014;2:74–7. https://doi.org/10.1016/j.mfglet.2014.01.005.

[21]  Belikovetsky S, Solewicz Y, Yampolskiy M, Toh J, Elovici Y, Gatlin J, et al. dr0wned - Cyber-Physical Attack with Additive Manufacturing. 11th USENIX Work. Offensive Technol. WOOT 2017, co-located with USENIX Secur. 2017, 2017.

[22]    Graves LMG, King W, Carrion P, Shao S, Shamsaei N, Yampolskiy M. Sabotaging metal additive manufacturing: Powder delivery system manipulation and material-dependent effects. Addit Manuf 2021:102029.

[23]    BSI. Die Lage der IT-Sicherheit in Deutschland 2014. 2014. https://doi.org/10.1021/j100331a045.

[24]    Hackers could destroy 3D printers by setting them on fire | TechRadar 2020. https://www.techradar.com/news/hackers-could-destroy-3d-printers-by-setting-them-on-fire (accessed February 23, 2023).

[25]    Renault-Nissan resumes nearly all production after cyber attack | Reuters 2017. https://www.reuters.com/article/us-cyber-attack-renault/renault-nissan-resumes-nearly-all-production-after-cyber-attack-idUSKCN18B0S5 (accessed February 23, 2023).

[26]    Honda's global operations hit by cyber-attack - BBC News 2020. https://www.bbc.com/news/technology-52982427 (accessed February 11, 2023).

[27]    Toyota cyberattack: Production to restart in Japan after attack on Kojima Industries | CNN Business 2022. https://www.cnn.com/2022/03/01/business/toyota-japan-cyberattack-production-restarts-intl-hnk/index.html (accessed January 19, 2023).

[28]    Colonial Pipeline Cyber Attack: Hackers Used Compromised Password - Bloomberg 2021. https://www.bloomberg.com/news/articles/2021-06-04/hackers-breached-colonial-pipeline-using-compromised-password (accessed January 27, 2023).

[29]    Rahman MH, Son Y-J, Shafae M. Graph-Theoretic Approach for Manufacturing Cybersecurity Risk Modeling and Assessment. ArXiv Prepr ArXiv230107305 2023.

[30]    Yampolskiy M, King WE, Gatlin J, Belikovetsky S, Brown A, Skjellum A, et al. Security of additive manufacturing: Attack taxonomy and survey. Addit Manuf 2018;21:431–57. https://doi.org/10.1016/j.addma.2018.03.015.

[31]    Wu M, Moon YB. Taxonomy of Cross-Domain Attacks on CyberManufacturing System. Procedia Comput. Sci., vol. 114, Elsevier B.V.; 2017, p. 367–74. https://doi.org/10.1016/j.procs.2017.09.050.

[32]    Pan Y, White J, Schmidt DC, Elhabashy A, Sturm L, Camelio J, et al. Taxonomies for Reasoning About Cyber-physical Attacks in IoT-based Manufacturing Systems. Int J Interact Multimed Artif Intell 2017;4:45–54. https://doi.org/10.9781/ijimai.2017.437.

[33]    Wu M, Moon YB. Taxonomy for secure cybermanufacturing systems. ASME Int Mech Eng Congr Expo Proc 2018;2:1–10. https://doi.org/10.1115/IMECE2018-86091.

[34]    Tuptuk N, Hailes S. Security of smart manufacturing systems. J Manuf Syst 2018;47:93–106. https://doi.org/10.1016/j.jmsy.2018.04.007.

[35]    Wu D, Ren A, Zhang W, Fan F, Liu P, Fu X, et al. Cybersecurity for digital manufacturing. J Manuf Syst 2018;48:3–12. https://doi.org/10.1016/j.jmsy.2018.03.006.

[36]    Elhabashy AE, Wells LJ, Camelio JA, Woodall WH. A cyber-physical attack taxonomy for production systems: a quality control perspective. J Intell Manuf 2019;30:2489–504. https://doi.org/10.1007/s10845-018-1408-9.

[37]    Mahesh P, Tiwari A, Jin C, Kumar PR, Reddy ALN, Bukkapatanam STS, et al. A Survey of Cybersecurity of Digital Manufacturing. Proc IEEE 2021;109:495–516. https://doi.org/10.1109/JPROC.2020.3032074.

[38]    Williams B, Soulet M, Siraj A. A Taxonomy of Cyber Attacks in Smart Manufacturing Systems. 6th EAI Int. Conf. Manag. Manuf. Syst., Springer; 2023, p. 77–97. https://doi.org/10.1007/978-3-030-96314-9_6.

[39]    Common Attack Pattern Enumeration and Classification (CAPEC) 2023. https://capec.mitre.org/ (accessed January 28, 2023).

[40]    Wu M, Zhou H, Lin LL, Silva B, Song Z, Cheung J, et al. Detecting attacks in cybermanufacturing systems: additive manufacturing example. Int. Conf. Mech. Mater. Manuf., vol. 108, EDP Sciences; 2017, p. 6005.

[41]    Sturm LD, Williams CB, Camelio JA, White J, Parker R. Cyber-physical vulnerabilities in additive manufacturing systems. Context 2014;7.

[42]    Stouffer K, Zimmerman T, Tang C, Lubell J, Cichonski J, Mccarthy J. NISTIR 8183 Revision 1, Cybersecurity Framework: Manufacturing Profile. 2020. https://doi.org/10.6028/NIST.IR.8183.

[43]    Blank RM, Gallagher PD. Guide for conducting risk assessments. 2012. https://doi.org/10.6028/NIST.SP.800-30r1.

[44]    Sailio M, Latvala O-M, Szanto A. Cyber threat actors for the factory of the future. Appl Sci 2020;10:4334. https://doi.org/10.3390/app10124334.

[45]    Automating threat actor tracking: Understanding attacker behavior for intelligence and contextual alerting - Microsoft Security Blog 2021. https://www.microsoft.com/en-us/security/blog/2021/04/01/automating-threat-actor-tracking-understanding-attacker-behavior-for-intelligence-and-contextual-alerting/ (accessed February 26, 2023).

[46]    Manufacturing Threat Perspective | Dragos 2020. https://www.dragos.com/resource/manufacturing-threat-perspective/ (accessed January 17, 2023).

[47]    Elhabashy AE, Wells LJ, Camelio JA. Cyber-physical attack vulnerabilities in manufacturing quality control tools.

Qual Eng 2020;32:676–92. https://doi.org/https://doi.org/10.1080/08982112.2020.1737115.

[48]     CVE - Home 2023. https://cve.mitre.org/cve/ (accessed February 11, 2023).

[49]     CWE - Common Weakness Enumeration 2023. https://cwe.mitre.org/ (accessed February 12, 2023).

[50]     Oltramari A, Cranor LF, Walls RJ, McDaniel PD. Building an Ontology of Cyber Security. STIDS, Citeseer; 2014, p. 54–61.

[51]     Obrst L, Chase P, Markeloff R. Developing an Ontology of the Cyber Security Domain. STIDS, 2012, p. 49–56.

[52]     Krauß D, Thomalla C. Ontology-based detection of cyber-attacks to SCADA-systems in critical infrastructures. 2016 Sixth Int. Conf. Digit. Inf. Commun. Technol. its Appl., IEEE; 2016, p. 70–3.

[53]     Tanner R, Loh NK. A taxonomy of multi-sensor fusion. J Manuf Syst 1992;11:314–25.

[54]     Longo F, Mirabelli G, Nicoletti L, Solina V. An ontology-based, general-purpose and Industry 4.0-ready architecture for supporting the smart operator (Part I–Mixed reality case). J Manuf Syst 2022;64:594–612.

[55]     Esmaeilian B, Behdad S, Wang B. The evolution and future of manufacturing: A review. J Manuf Syst 2016;39:79–100. https://doi.org/https://doi.org/10.1016/j.jmsy.2016.03.001.

[56]     Framework for Improving Critical Infrastructure Cybersecurity, Version 1.1. 2018. https://doi.org/10.6028/NIST.CSWP.04162018.

[57]     Kaspersky. The Human Factor in IT Security: How Employees are Making Businesses Vulnerable from Within. 2022. https://www.kaspersky.com/blog/the-human-factor-in-it-security/ (accessed February 3, 2023).

[58]     Symantec. Internet Security Threat Report VOLUME 24, February 2019. vol. 24. 2019.